\documentstyle[12pt,fleqn,cite]{article}

\def\be{\begin{equation}}
\def\ee{\end{equation}}
\def\bea{\begin{eqnarray}}
\def\eea{\end{eqnarray}}

\begin{document}
\begin{titlepage}
\begin{center}
\hfill hep-th/0104122\\
\hfill IITK/PHY/2001/7\\
\vskip .2in

{\Large \bf Rotating Brane World Black Holes}
\vskip .5in

{\bf Moninder Singh Modgil, Sukanta Panda and Gautam Sengupta}\\
\vskip .1in
{\em Department of Physics,\\
Indian Institute of Technology\\
Kanpur 208 016\\
INDIA}
\vskip .5cm
\end{center}

\begin{center} {\bf ABSTRACT}
\end{center}
\begin{quotation}\noindent
\baselineskip 10pt

A five dimensional rotating black string in a Randall-Sundrum brane world
is considered. The black string intercepts the three brane in a four
dimensional rotating black hole. The geodesic equations and the
asymptotics in this background are discussed.

\end{quotation}
\vskip .2in
April 2001\\
\end{titlepage}
\vfill
\eject

Unification of gravity with the other fundamental interactions in the
context of string theory have suggested the possibility of our universe
being higher dimensional. The extra spatial dimensions in these scenarios
are assumed to be compactified and hence unobservable at low energies.
Recent advances have however shown that if the matter fields are localized
on a three-brane (or a smooth domain wall) and the electroweak scale
assumed to be the fundamental scale then the extra dimensions are not
restricted to be small \cite{add}.  This removes the hierarchy between the
electroweak scale and the Planck scale but introduces a new hierarchy of
scales amongst the large and the small dimensions. This construction has
opened the way for interesting phenomenological possibilties of low scale
quantum gravity effects which may be accessible in the next generation
accelarators. Recently Randall and Sundrum (RS) \cite{rs12} considered a
five dimensional modified version of these models in which the matter
fields were restricted on a three brane with the brane coordinates being
dependent on the location in the fifth dimension. These compactification
geometries have been termed {\it warped compactification}. The models also
require a regulator three brane at a certain distance in the fifth
dimension. For this construction to be an acceptable solution of the five
dimensional Einstein equations the three brane(s) is required to be
embedded in a slice of five dimensional Anti de-Sitter(AdS) space-time.
Interestingly in this model four dimensional gravity arises due to the
localized zero-mode of the five dimensional Kaluza-Klein graviton. This
leads to an interesting resolution of the hierarchy problem as the extra
fifth dimension is not required to be large in this case. It could be
shown later that the regulator brane in this scenario could be removed to
infinity leading to a model with effectively a single three brane and an
infintely large extra fifth dimension. The insights gained from these
constructions have resulted in intense activity in this area in the recent
past. A full non linear treatment in the framework of supergravity
\cite{gibs} have confirmed the conclusions resulting from the linearized
approximation.  In particular, these constructions have led to the
possibility of detecting low scale quantum gravity effects on phenomenlogy
at the weak scale in the next generation particle acclerators \cite{lyk}.  
The cosmological consequences of these and related models are also being
pursued.

A related direction of investigation has been the study of four
dimensional black holes on the three-brane from this {\it brane world}
perspective. In an interesting article Chamblin, Hawking and Reall(CHR)
\cite {chr} have constructed a Schwarzschild black hole on the four
dimensional world volume of the three brane from the five dimensional RS
scenario. Such a configuration is extended in the extra dimensions and
will be a higher dimensional object in the brane world. such a solution
would presumably describe the end state of gravitational collapse of
non-rotating matter on the three brane. The metric on the brane in this
case is required to be a Schwarzschild metric to satisfy the usual
astrophysical observations and the requirements of the singularity
theorems. However, the obvious choice of an AdS-Schwarzschild solution in
five dimensions fails to satisfy the Israel junction conditions at the
location of the brane which are required to be $Z_2$ reflection symmetric.
CHR considered an alternative configuration of a five dimensional black
string which intercepted the three brane in a four dimensional
Schwarzschild metric. This has the usual Schwarzschild singularity on the
brane and is extended along the transverse direction. However it is
observed that the solution is singular at the AdS horizon far away from
the brane. A study of the geodesics show the that this singularity is
actualy a parallely propagated (p-p) curvature singularity \cite
{pps,gibs}. The solution is unstable far away from the brane due to the
Gregory-Laflamme instability \cite{greg} and is conjectured to evolve to a
black cigar solution which looks like a black string far away from the AdS
horizon but closes off before reaching it. The exact metric for this cigar
solution is yet to be determined. It was however pointed out in \cite
{greg1} through an explicit calculation that in these asymptoticaly AdS
solutions there would be an accumulation of mini black holes towards the
AdS horizon which do not indicate either a cigar or a pancake geometry in
the bulk. Attempts to consider the off-brane metric in the linearized
framework have also appeared in \cite {gian}. Investigation of similar
constructions in a 3 + 1 dimensional RS model with two branes, where exact
solutions in the form of AdS C metrics are available, show that the
corresponding black hole solutions are non-singular at the AdS horizon
\cite{emp}. Generalizations of this construction for higher dimensional
brane worlds \cite{kim} and charged black holes in the RS models have been
obtained \cite {cbh}. Numerical studies for the off brane cigar metrics
have also been performed \cite {num}.

In this brief report we construct a four dimensional Kerr black hole on
the three-brane in a five dimensional RS brane world scenario.  It is
apparent that a AdS-Kerr solution in five dimensions will fail to satisfy
the junction conditions at the three-brane.
The reason being that the AdS-Kerr solution reduces to the
AdS-Schwarzschild solution for the rotation parameter going to zero, which
fails to satisfy the junction conditions appropriate to a vaccuum solution
\cite {chr}. Thus, we consider a rotating black string in five dimensions
which can be shown to intercept the three brane in a four dimensional Kerr
metric. We should mention here that the usual rotating black holes in the
brane world and the BTZ variants have also been obtained in \cite{emp} for
two branes embedded in a (3+1) dimensional AdS space-time where exact AdS
C metrics are known. However such a construction is unavailable in five or
higher dimensions where exact solutions are not apparent and it is
neceesary to work in the framework of linearized gravity. This
construction also serves as a prelude to obtaining the four dimensional
Kerr-Newman solutions in a brane world scenario. Such higher dimensional
constructions will also be relevant to general string compactifications.
We find that like the Schwarzschild case the Kerr solutions are also
singular at the AdS horizon as expected in the linearized approximation.
Study of the geodesic equations clearly illustrate that the singularity at
the AdS horizon is accessible only on bound state orbits. This indicates
the existence of parallely propagated (pp) curvature singularities at the
AdS horizon in this metric also.

The Randall-Sundrum models are based on a five dimensional AdS as the
bulk space-time with the metric
\be
ds^2={l^2\over {z^2}}(\eta _{ij}dx^idx^j + dz^2)
\ee
where $z=0$ and $ z=\infty$ are the conformal infinity and the AdS
horizon respectively. Here $(i,j)$ runs over the four dimensional world
volume of the
three brane and $l$ is an AdS length scale.
The actual RS geometry is obtained by slicing off the small $z$ region at
$z=z_0$ and glueing a copy of the large $z$ geometry. The
resulting topology is
essentialy $R^4 \times {{S^1}\over {Z_2}}$.
The discontinuity 
of the extrinsic curvature at
the $z=z_0$ surface correponds to a thin distributional source of stress-
energy. From the Israel junctions conditions this may be interpreted as
a relativistic three brane with a corresponding tension \cite{gid}.
Another variant of this model is to slice the AdS space-time
both at $z=0$ and $z=l$ and insert three branes with $Z_2$ reflection
symmetry at both the surfaces. The Israel conditions now require a
negative tension for the brane at $z=l$. The first version may be obtained
from the second by allowing the negative tension brane to approach the
AdS horizon at $z=\infty$ however a dynamical realization of this is not
yet clear. We will focus our considerations on the first
variant of the RS geometry in this case but state that our construction
should be trivialy generalized to the second variant also.
The interesting consequence of the RS construction is that perturbations
of
the five dimensional metric are normalizable modes peaked at the location
of the brane. In particular the zero mode satisfying the vaccuum Einstein
equation yields standard four dimensional gravity on the brane. 

The Einstein equations in five dimensions with a negative cosmological
constant continue to be satisfied for any metric $g_{ij}$ which is Ricci
flat. CHR \cite{chr} considered the gravitational collpase of non-rotating
matter trapped on the brane to form a black hole. The metric on the brane
in this case must be a Schwarzschild metric which is Ricci flat. However
the obvious choice of a five-dimensional AdS-Schwarzschild metric fails to
satisfy the Israel junction conditions compatible with the $Z_2$
reflection symmetry. CHR \cite{chr} took the natural choice of the four
dimensional metric in eqn (1) to be Schwarzschild, thus arriving at the
metric which describes a five dimensional uncharged black string, in the
RS geometry. The black string metric is given by, \begin{eqnarray} ds^{2}
& = & \frac{l^2}{z^2} [-U(r) dt^2 + U^{-1}(r) dr^2 + r^2 (d\theta^2 +
sin^2{\theta} d\phi^2) + dz^2 ], \end{eqnarray} where $U(r)=1-{2M\over
r}$. Inclusion of a three brane with reflection symmetry now requires the
brane tension to be compatible with the junction conditions relating the
extrinsic curvatures on both sides. This gives \begin{equation} T_3=\pm
\frac{6}{\kappa^2l}. \end{equation} where $\kappa^2=8\pi G_5$ where $G_5$
is the five dimensional Newton's constant. The single brane model however
will correspond to a positive brane tension. For the brane at $z=z_0$ one
introduces the coordinate $w=z-z_0$ and the metric is then given as \bea
ds^2 &=& {l^2\over {(\mid w \mid + z_0)^2}}\big[ -U(r)dt^2
+U(r)^{-1}dr^2\\ & & +r^2(d\theta^2 + sin^2\theta d\phi^2) + dw^2\big]
\eea with $-\infty < w < \infty$ and the brane at $w=0$. The metric on the
brane may be recast into the standard Schwarzschild form by rescaling the
$t$ and $r$ coordinates. The ADM mass of the black hole to an observer on
the brane is then ${\tilde M}=M{l\over {z_0}}$ and the proper horizon
radius is $2{\tilde M}$. The square of the curvature scalar is given as
\be R_{\mu\nu\rho\sigma}R^{\mu\nu\rho\sigma}={1\over {l^4}}\bigg(40 - {48
M^2z^4\over r^6}\bigg), \ee where $\mu, \nu$ goes over the five
dimensions. This diverges at the AdS horizon $z=\infty$ and posesses the
usual black string singularity at $r=0$. However an examination of the
geodesic equations in this space-time show that curvature invariants
diverge only for the bound state geodesics that encounters the $r=0$
singularity but are finite along non-bound state geodesics which reach
$r=\infty$. In \cite{chr} CHR examines the curvature components in an
orthonormal frame parallely propagated along a non-bound timelike
geodesic. It is observed that some of the curvature components in this
frame diverges at $z=\infty$. This signals the presence of parallely
propagated curvature singularities \cite {wald, pps, gibs} arising out of
tidal effects at the AdS horizon. CHR further argue that the black string
metric is unstable due to the standard Gregory-Laflamme instabilty near
the AdS horizon where the black string behaves as if it is in an
asymptoticaly flat space-time provided the horizon radius is small. But
the AdS space acts like a confining box preventing development of
perturbations with wavelengths larger than the AdS length scale so
instabilities must occur at shorter wavelengths hence at large values of
$z$. This insatbility may cause the black string horizon to pinch off
before the AdS horizon leading to a bulk metric describing a black cigar.
The metric on the brane being a stable Schwarzschild metric far away from
the AdS horizon. Careful consideration of the black string instability in
AdS spaces \cite {greg1} however seems to indicate an accumulation of mini
black holes towards the AdS horizon.

We generalize the construction of CHR\cite {chr} to study the occurrence
of a rotating uncharged black hole on the three brane. Four dimensional
General Relativity on the brane requires this to be the Kerr metric which
is also Ricci flat. The extension to a five dimensional bulk AdS-Kerr
metric fails to satisfy the junction conditions at the location of the
brane consistent with a vaccuum configuration. This is because in the
absence of rotation the AdS-Kerr solution will reduce to the
AdS-Schwarzschild metric which has been shown to be incompatible with the
junction conditions in \cite {chr}. This leads to our choice of the bulk
metric as a rotating black string in five dimensions which is consistent
with the junction conditions. The metric for the brane world black string
in the Boyer-Lindquist coordinates is given as, \begin{eqnarray} ds^{2} &
= & \frac{l^{2}}{(z_{0}+|w|)^2}\left[-\big(\frac{\Delta
-a^2sin^2\theta}{\Sigma}\big)dt^{2} +\frac{\Sigma} {\Delta} dr^2 \right.  
\nonumber \\
 & + &\left.  \Sigma d\theta^{2} + \frac{4aMr\sin^2{\theta}}
{\Sigma}d\phi  dt \right. \nonumber \\
& + & \left. \big[\frac {(r^2 + a^2)^2 -\Delta a^2
sin^2\theta}{\Sigma}\big]
sin^2\theta d\phi^2
+ dw^2\right].
\end{eqnarray}
Where we have
\bea
\Sigma=r^2 + a^2 cos^2\theta
\eea
and
\bea
\Delta=r^2 + a^2 -2 Mr.
\eea
Here the location of the brane is at $z=z_0$  with the $Z_2$ reflection
symmetry imposed there.
Notice that the four dimensional part of the metric is Ricci flat apart
from the conformal factor and as
a consequence this guarantees the satisfaction of the five dimensional
Einstein equations with a negative cosmological constant.
As in the case of CHR \cite {chr} the induced metric on the brane may be
recast in the form of the standard four dimensional Kerr-metric by
suitable rescaling. The ADM mass and the angular momentum for the rotating
black hole on the brane are then given as $M^*={Ml\over {z_0}}$ and
$J^*={l^2\over {z_0^2}}Ma$ to an observer confined to the brane. We
assume here that $a^2\leq M^2$ to avoid the occurrence of naked
singularities \cite{wald}. The Kerr metric on the brane will exhibit
the usual features of the inner and outer horizons and an ergosphere. 
The horizons will now be given
for $r{\pm}=M^* \pm (M^{*2} - a^{*2})^{1\over 2}$ and the stationary
limit surface at $r_s=M^* + (M^{*2} - a^{*2}cos^2\theta)^{1\over
2}$.

The square of the curvature tensor is determined as \cite{emp},
\begin{equation}
R_{\mu\nu\rho\sigma}R^{\mu\nu\rho\sigma}=\frac{1}{l^4}\big [40
+\frac{48M^2z^4}{\Sigma^6}(r^2-a^2 cos^2{\theta})(r^4-14a^2r^2cos{\theta}
+a^4cos^4{\theta})\big ] \end{equation} This shows the usual ring
singularity of the Kerr metric for $\rho=0$ and $\theta={\pi\over 2}$ and
diverges at the AdS horizon at $z=\infty$. The ring singularity for the
rotating black string has a translational symmetry in the transverse
direction and is thus an example of an extended singular solution in five
dimensions which is asymptoticaly AdS.

Our metric for the rotating black string being a stationary axisymmetric
metric posesses two timelike Killing isometries. If $u$ is the tangent
vector to a timelike or null geodesic with an affine parameter $\lambda$,
the timelike Killing vectors $\xi={\partial \over{\partial t}}$ and
$\chi={\partial \over{\partial t}}$ gives rise to two conserved quantities
$E=-\xi.u$ and $L=\chi.u$. Rearrangement of these equations \cite{wald}
provides us with the geodesic equations for the $t$ and $\phi$ directions
for motion in the equatorial plane $\theta={\pi\over 2}$. These turn out
to be as follows
\begin{equation} \frac{dt}{d\lambda}= \frac{z^2}{l^2\Delta}\left[
\left(r^2+a^2+\frac{2a^2M}{r}\right) E-\frac{2aM}{r}L\right]
\end{equation} \begin{equation}
\frac{d\phi}{d\lambda}=\frac{z^2}{l^2\Delta}\left[\left(1-\frac{2M}{r}\right)L
+\frac{2aM}{r}E\right] \end{equation} The $z$ equation is then given as
\begin{equation}
\frac{d}{d\lambda}\left(\frac{1}{z^2}\frac{dz}{d\lambda}\right)=-
\frac{\sigma} {zl^2}. \end{equation} Here $\sigma=0$ for null and
$\sigma=1$ for the timelike geodesics. The solutions for null geodesics
are $z=$constant or \begin{equation} z=\frac{-z_{1}l}{\lambda},
\end{equation} The solution for a timelike geodesic is \begin{equation}
z=-z_{1}cosec(\lambda/l). \end{equation} The solution $z=const. $ relates
to the Schwarzschild case hence we focus our attention on the other
solutions which appear to reach the AdS horizon at $z=\infty$. For this
class of solutions the radial geodesic equation in the equatorial plane is
computed to be \begin{equation}
\left(\frac{dr}{d\lambda}\right)^2+\frac{z^4}{l^4}\left[\left(\frac{L^2-a^2E^2}
{r^2}-\frac{2M}{r^3}(aE-L)^2-E^2\right) \right ]+\frac{l^2}{z_{1}^2}
\frac{\Delta}{r^2} =0 \end{equation} Following CHR we now introduce a new
parameter $\nu=-{z_{1}^2\over \lambda}$ for the null geodesics and
$\nu=-(z_{1}^2/l)cot({\lambda\over l})$ for the timelike geodesics. We
also define new coordinates $\tilde{r}=z_{1}r/l,\tilde{t}=z_{1}t/l$, and
new constants $\tilde {E}=z_{1}E/l,\tilde{L}=z_{1}^2L/l^2,
\tilde{M}=z_{1}M/l$ and $\tilde{a}=z_{1}a/l$. Using these rescaled
qunatities in the radial equation it is possible to remove the explicit
$z$ dependence and the equation may be recast in an equivalent four
dimensional form. This matches with the radial equation for a timelike
geodesic in a four dimensional Kerr black hole with a mass ${\tilde M}$
\cite{wald}.
\begin{equation}
\left(\frac{d\tilde{r}}{d\nu}\right)^2 + \left[\frac{\tilde{L}^2
-\tilde{a}^2\tilde{E}^2}{\tilde{r}^2}-\frac{2\tilde{M}}{\tilde{r}^3}
(\tilde{a}\tilde{E}-L)^2-\tilde{E}^2+\frac{\tilde{\Delta}}{\tilde{r}^2}\right]=0
\end{equation}
In this case $\nu$ is the proper time along the geodesic. Notice that
both the timelike and null geodesic in five dimensions gives rise to
timelike four dimensional geodesics. 

It is now possible to study the behaviour near the singularity that is
$\lambda \rightarrow 0^-$ which is equivalent to the four dimensional
affine parameter $\nu\rightarrow \infty$. This describes the late time
behaviour of the corresponding four dimensional geodesics. The treatment
is simmilar to CHR \cite {chr}. The geodesics which reach the singularity
at ${\tilde r}=0$ do so in finite affine parameter $\nu$. For infinite
$\nu$ there are two cases. For one the geodesic reaches ${\tilde
r}=\infty$ and for the other we have bound states or orbits restricted to
finite ${\tilde r}$ outside the horizon.  It is evident from the
expression eqn.(8) for the singularity that for the orbits that reach
$r=\infty$ or non bound state orbits, the curvature squared remains finite
at $z\rightarrow \infty$, the AdS horizon. However, the bound state orbits
at finite $r$ encounter a curvature singularity at the AdS horizon. This
is completely similar to the Schwarzschild case as in CHR where the
singularity at the AdS horizon is a parallely propagated (p-p) curvature
singularity. It is conceivable that in the rotating case too the
singularity at the AdS horizon is also a pp singularity.

To determine this it is necessary to calculate the projection of the
curvature tensor to an orthonormal frame which is parallely propagated on
a non bound timelike geodesic. In the case of the non-rotating
Schwarzschild metric this reduces to the problem of determining a unit
normal to the tangent to a time like non bound geodesic. For this diagonal
metric this becomes an effectively two dimensional problem for a specific
choice of the normal. However the non-diagonal Kerr metric such a
determination of the corresponding normal reduces to an effectively three
dimensional exercise. This requires the solution of three coupled partial
differential equations in order to satisfy the parallel transport
equations and becomes computationaly intractable. However we emphasize
that such an orthonormal frame should exist in which it would be possible
to show that the singularity at the AdS horizon is indeed a pp curvature
singularity.

Following CHR \cite{chr} it is conceivable that the rotating black string
will also be subject to the Gregory-Laflamme instability \cite{greg}
arising out of long wavelength perturbations. This would make the black
string unstable near the AdS horizon but stable far away causing the
horizon to pinch off and form a line of mini black holes. However from the
considerations in \cite{greg1} for Schwarschild black strings in AdS space
this instability would tend to form an accumulation of mini black holes
towards the AdS horizon. As shown in \cite{greg1} this would result in a
"sttotie" shape for the event horizon of a Schwarschild black string with
masses comparable to the AdS curvature. For masses much larger than the
AdS curvature the resulting solution will look like a five dimensional
black hole. The arguments in \cite{greg1} should generalize also for the
case of the rotating black string presented here, however careful
considerations are needed for the two horizons and the stationary limit
surface. Presumably the instabilty would also cause the black hole to be
unstable near the AdS horizon signaling the presence of the pp
singularity. As shown in \cite{emp} in the 3 + 1 dimensional RS model the
pp singularity at the AdS horizon seems to be an artifact of the
linearized approximation. This is also borne out in the full non linear
supergravity considerations in \cite {gibs}

In conclusion we have obtained a description of a four dimensional
uncharged rotating black hole on the brane described by a Kerr metric from
a five dimensional RS brane world perspective. The bulk five dimensional
solution which intercepts the three brane in a rotating black hole has
been considered to be a five dimensional rotating black string in the RS
geometry. This choice is compatible with the requirement of the Israel
junction conditions arising out of the reflection symmetry at the location
of the three brane in the RS brane world. The Kerr metric on the brane is
shown to exhibit the usual ring singularity in addition to a singularity
at the AdS horizon. Furthermore we have obtained the geodesic equations in
the equatorial plane for this background and observe that the singularity
at the AdS horizon is accessible only on bound time like geodesics which
remain at finite orbit radius. This signals the occurrence of parallely
propagated (pp) curvature singularity at the AdS horizon. The construction
of orthonormal frame to explicitly study the pp singularity seems
non-trivial due to the non diagonal nature of the extended Kerr metric for
the black string. We further state that the solution presented will be
subject to the usual black string instability causing the formation of
mini black holes which would accumulate towards the AdS horizon. However
an explicit calculation in the spirit of \cite {greg1} is necessary for
the rotating black string case considered here. It is also imperative to
construct an exact description of the off-brane bulk metric. This would
presumably be a generaliztion of the AdS C-metric in four dimensions. It
would be relevant to generalize our construction for the full Kerr-Newman
metric on the brane which would describe a charged rotating black hole in
a brane world. Work is in progress along these directions. \bigskip

{\bf Acknowledgements :} All of us would like to thank the members of the
High Energy Physics Group at IIT Kanpur for many discussions, especialy S.
Mukherji, S. D. Joglekar, V. Ravishankar and D Sahdev. GS would also like
to thank S. W. Hawking for discussion during Strings 2001 and K. Ray for
certain elucidations.

\vfil
\eject

\end{document}